\documentclass[cameraready]{Interspeech}

\usepackage{hyperref}
\usepackage{url}
\usepackage{booktabs}
\usepackage{amsfonts}       
\usepackage{nicefrac}       
\usepackage{xcolor}         
\usepackage{graphicx}
\usepackage{wrapfig}
\usepackage{amsmath}
\usepackage{bm}
\usepackage{multirow}
\usepackage[table]{xcolor}
\usepackage[title]{appendix}
\usepackage{float}
\usepackage{makecell}

\title{ELSA: Acoustic Event-Level Semantic Alignment for\\Fine-Grained Reference-Free Text-to-Audio Evaluation}

\author[affiliation={1}, orcid=0009-0008-5564-3835, equalcontribution]{Shuntaro}{Suzuki}
\author[affiliation={1}, orcid=0009-0009-4524-9022, equalcontribution]{Kento}{Tokura}
\author[affiliation={1}, orcid=0009-0005-2087-2038, equalcontribution]{Daichi}{Yashima}
\author[affiliation={1}, orcid=0009-0001-9655-7561, equalcontribution]{Kanon}{Amemiya}
\author[affiliation={1}, orcid=0000-0002-0261-0510]{\\Komei}{Sugiura}
\author[affiliation={1}, orcid=0000-0003-0520-7847]{Shinnosuke}{Takamichi}

\address{
    $^1$ Keio University, Japan
}
\email{\{shuntaro20021227, tkento1985, ydaichi1207, kanon-amemiya\}@keio.jp}
\keywords{text-to-audio, automatic evaluation metric, fine-grained semantic similarity, CLAPScore}

\usepackage{comment}

\begin{document}
\maketitle
\setlength{\abovedisplayskip}{3pt} 
\setlength{\belowdisplayskip}{3pt} 

\begin{abstract}
\vspace{-2mm}

Text-to-audio (TTA) generation, synthesizing audio from natural language, has been widely studied for its ability to capture precise user intent. 
To effectively advance TTA models, it is essential to reliably evaluate generated audio without relying on costly human subjective ratings, motivating the development of automatic evaluation metrics that correlate well with human judgments. 
While recent CLAP-based metrics provide practical reference-free solutions, their coarse-grained text–audio similarity matching often correlates poorly with human ratings. 
To address this, we propose ELSA, a reference-free evaluation metric for fine-grained text–audio alignment. 
ELSA decomposes generated audio guided by distinct acoustic events derived from the text query and assesses event-level alignment.
Experiments across four TTA benchmarks show that ELSA reveals a higher correlation with human subjective ratings than prior metrics, highlighting its effectiveness for reliable TTA evaluation.\footnote{Project page: \href{https://elsa-projectpage.pages.dev/}{https://elsa-projectpage.pages.dev/} }
\end{abstract}

\vspace{-2mm} 
\section{Introduction}
\vspace{-2mm} 

Audio generation conditioned on user intent, encompassing speech, sound effects, and music, has been widely studied~\cite{PixelSNAIL, DAG}.
This interest is motivated by diverse applications such as augmented reality audio environments and media sound generation~\cite{Lloyd, Riffusion}, with text-to-audio (TTA) generation gaining particular attention for directly synthesizing audio from text while accurately capturing user intent~\cite{AudioLDM2, AudioX, TangoFlux}.



Reliable evaluation of generated audio is essential for the development and benchmarking of TTA models.
Although human assessment of overall audio quality (OVL) and relevance to the input text (REL) is considered the gold standard~\cite{AudioGen}, it is costly and time-consuming, motivating the development of automatic metrics that correlate with human ratings~\cite{PAM, AudioBERTScore}.

Automatic evaluation metrics for TTA can be broadly categorized into reference-based metrics~\cite{SI-SDR, AudioBERTScore} and reference-free metrics~\cite{PAM, CLAPScore}, depending on whether the corresponding reference audio is required.
Reference-based metrics enable fine-grained evaluation by directly comparing audio signals. 
For example, AudioBERTScore~\cite{AudioBERTScore} computes frame-level similarity in the feature space of an audio foundation model, thereby enabling implicit assessment at the level of individual acoustic events.
However, the requirement for reference audio substantially limits their applicability in real-world scenarios.
In contrast, reference-free metrics compare the input text with the generated audio directly, making them applicable to a wider range of experimental settings, including real-world scenarios.
Nonetheless, because these methods compare fundamentally different modalities (i.e., text and audio), their evaluations are typically restricted to coarse-grained similarity measured in a jointly aligned text–audio embedding space, as shown by CLAP-based metrics~\cite{CLAPScore}.
As a result, their correlation with human subjective ratings remains limited. 
For instance, on the RELATE dataset, the Spearman correlation between CLAPScore and human subjective ratings of REL is only 0.280~\cite{HumanCLAP}.

\begin{figure}[t]
    \centering
    \includegraphics[width=0.94\linewidth]{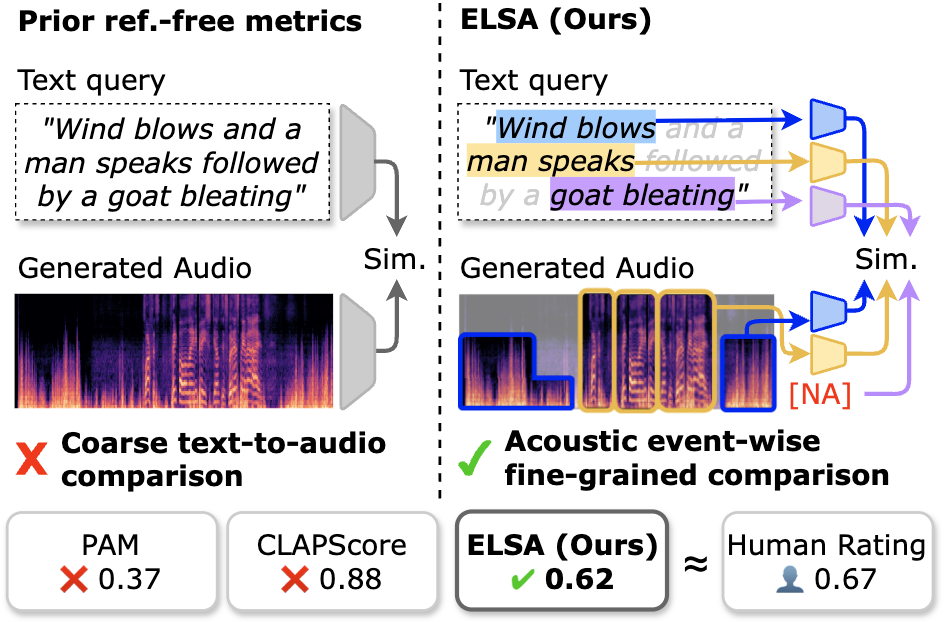}
    \vspace{-3mm} 
    \caption{Overview of ELSA. ELSA enables acoustic event-wise, fine-grained text–audio alignment for reference-free TTA evaluation, yielding high correlation with human subjective ratings.}
    \label{fig:eye-catch}
    \vspace{-2mm} 
\end{figure}

To this end, we propose ELSA\footnote{Acoustic \textbf{E}vent-\textbf{L}evel \textbf{S}emantic \textbf{A}lignment score.}, a reference-free evaluation metric for TTA generation that enables fine-grained text–audio comparison.
Figure~\ref{fig:eye-catch} presents the core idea.
Unlike existing methods that rely on global text–audio matching, ELSA decomposes the text query into semantically distinct acoustic events and performs event-wise alignment with the corresponding audio segments.
This design allows for a more precise assessment of short and transient acoustic events (e.g., footsteps) that are often overlooked by coarse-grained metrics~\cite{CLAPScore}.
Experimental results demonstrate that ELSA consistently achieves higher correlation with human ratings than existing metrics across four TTA benchmarks.
In addition, ablation and sensitivity analyses demonstrate the contribution of individual components and the robustness of ELSA to the number of target acoustic events.

\begin{figure*}[t!]
    \centering
    \includegraphics[width=0.87\linewidth]{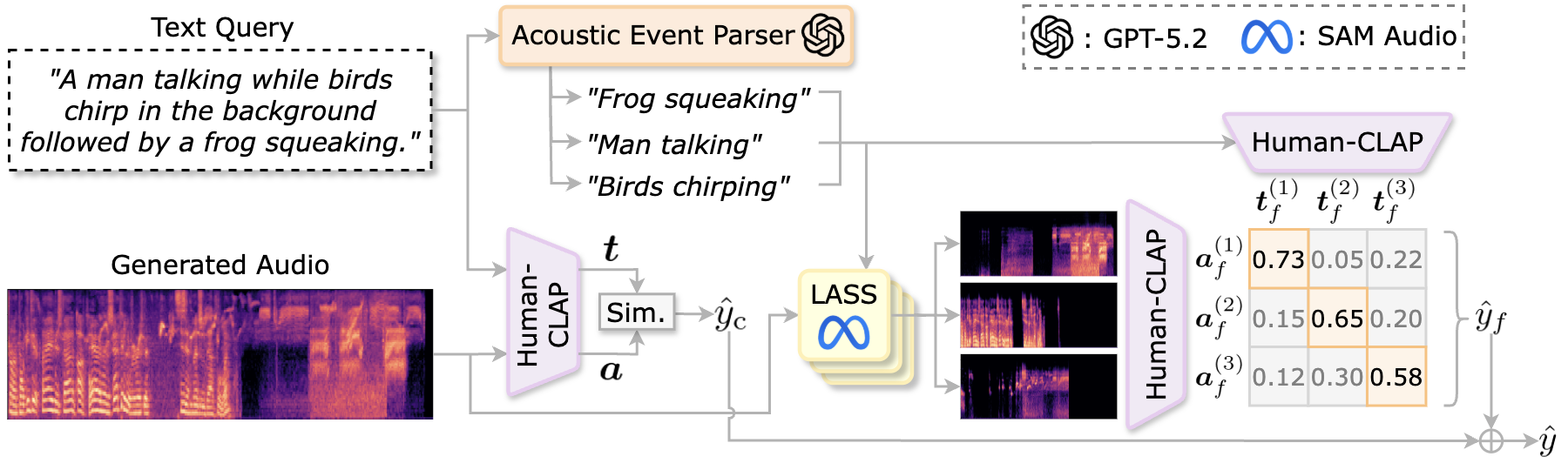}
    \vspace{-3mm} 
    \caption{Architecture of ELSA. ELSA hierarchically evaluates global text–audio matching and fine-grained acoustic-event alignment by combining shared text–audio embeddings with event-level representations extracted via a text parser and a language-queried audio source separation (LASS) model.
    }
    \vspace{-2mm} 
    \label{fig:model_arch}
\end{figure*}

\vspace{-2mm}
\section{Related Works}
\vspace{-2mm}

Automatic evaluation metrics for TTA generation~\cite{SDR, SI-SDR} have been comprehensively reviewed in Lan et al.~\cite{lan} and Su et al.~\cite{su}.
Among these, reference-free metrics such as PAM~\cite{PAM} and CLAPScore~\cite{CLAPScore} have gained prominence due to their broad applicability, avoiding the need for reference audio by directly aligning text and audio representations within a shared CLAP~\cite{MS-CLAP} embedding space.
Moreover, Human-CLAP~\cite{HumanCLAP} improves upon this framework by refining the CLAP feature space itself to better reflect human perceptual judgments, thereby providing a more reliable evaluation measure.

Conversely, cross-modal evaluation in the vision–language domain~\cite{vela, EMScore, HICEScore} has advanced significantly by decomposing inputs into semantically coherent units for fine-grained alignment, thereby achieving stronger correlations with human ratings.
Although several audio–language studies have explored multi-granular representation learning~\cite{MGA-CLAP, FineLAP}, existing reference-free TTA evaluation metrics still primarily rely on holistic text–audio representations and lack explicit semantic decomposition for fine-grained evaluation.

\vspace{-2mm} 
\section{Method}
\vspace{-2mm} 

An overview of the proposed metric is shown in Figure~\ref{fig:model_arch}.
ELSA is inspired by prior evaluation metrics for video and image captioning~\cite{vela,pearl}, particularly those that hierarchically assess visual–language alignment~\cite{HICEScore, EMScore}.

\vspace{-2mm} 
\subsection{Event-Level Representation Extraction}
\vspace{-2mm} 
Fine-grained text–audio comparison is achieved by hierarchically extracting user intent from the text query, as well as various acoustic characteristics present in generated audio.
Given a text query $\bm{x} \in \mathbb{R}^{V \times N}$ and the corresponding generated audio $\bm{s} \in \mathbb{R}^{L}$, we first embed both modalities into a shared text–audio feature space using a pretrained embedder.
This yields global text and audio representations, $\bm{t} \in \mathbb{R}^{d}$ and $\bm{a} \in \mathbb{R}^{d}$, respectively.
Here, $V$, $N$, $L$, and $d$ denote the vocabulary size, maximum token length, audio sequence length, and embedding dimension, respectively.
Throughout our framework, we employ Human-CLAP~\cite{HumanCLAP}\footnote{https://github.com/sarulab-speech/Human-CLAP}
 as the embedder, as it demonstrated the most favorable behavior in our preliminary experiments among several candidate text–audio embedding models.
These global representations provide a coarse-grained reflection of user intent.
However, because the global embeddings aggregate semantics over the entire sequence, transient acoustic events can contribute less to the representation and may be obscured.

To address this limitation and better reflect fine-grained user intent specified in the text query, we additionally extract acoustic event-level representations. 
Specifically, we decompose the text query $\bm{x}$ into a set of semantically distinct events $\{\bm{x}_f^{(i)}\}_{i=1}^{M}$using a text-only LLM. 
For each event description $\bm{x}_f^{(i)}$, we obtain a corresponding text representation $\bm{t}_f^{(i)}$. 
Furthermore, we employ a Language-queried Audio Source Separation (LASS) model to identify an event-relevant audio segment and extract its audio representation $\bm{a}_f^{(i)}$ conditioned on $\bm{x}_f^{(i)}$. 
In our implementation, we use GPT-5.2\footnote{https://openai.com/index/introducing-gpt-5-2/}  as the LLM and SAM Audio~\cite{SAM-Audio}\footnote{https://github.com/facebookresearch/sam-audio} as the LASS model. 
To ensure compatibility with the LASS training distribution, each $\bm{x}_f^{(i)}$ is constrained to a concise noun–verb phrase (e.g., “dog barking.”)

\begin{table*}[t]
    \centering
    \caption{Correlation with human subjective ratings across baseline metrics.
    $\rho$ and $\tau$ denote Spearman’s $\rho$ coefficient and Kendall’s $\tau$ coefficient, respectively.
    \textbf{Bold} and \underline{underlined} values indicate the best and second-best results.}
    \vspace{-3mm} 
    \label{tab:quantitative_main} 
    \resizebox{\textwidth}{!}{
    \begin{tabular}{l c c c c c c c c c c c c c c}
      \toprule
      \multirow{3}{*}{Metrics [\%]}
      &\multicolumn{4}{c}{AudioCaps}
      &\multicolumn{4}{c}{Clotho}
      &\multicolumn{4}{c}{MusicCaps}
      &\multicolumn{2}{c}{RELATE}\\
      \cmidrule(l{1mm}r{1mm}){2-5}
      \cmidrule(l{1mm}r{1mm}){6-9}
      \cmidrule(l{1mm}r{1mm}){10-13}
      \cmidrule(l{1mm}r{1mm}){14-15}
      &\multicolumn{2}{c}{OVL}& \multicolumn{2}{c}{REL}
      &\multicolumn{2}{c}{OVL}& \multicolumn{2}{c}{REL}
      &\multicolumn{2}{c}{OVL}& \multicolumn{2}{c}{REL}
      &\multicolumn{2}{c}{REL}\\
      &$\rho$& $\tau$&$\rho$& $\tau$
      &$\rho$& $\tau$&$\rho$& $\tau$
      &$\rho$& $\tau$&$\rho$& $\tau$
      &$\rho$& $\tau$\\
      \midrule
      \rowcolor{gray!10}
      \multicolumn{15}{l}{\textit{Reference-based Metrics}}\\
      SI-SDR
      & -2.8& -2.4& -2.9& -2.4
      & -4.4& -3.7& 0.2& 0.2
      & 3.7& 3.0& 4.8& 4.0
      & -2.4& -2.0\\
      FD$_{\text{OpenL3}}$
      &-6.9 &-4.5 &-3.1 &-1.9
      & -5.1& -3.6& 4.7& 3.2
      & 11.7 & 8.0& 5.9& 3.8
      & -2.1& -1.5\\
      KL$_{\text{PaSST}}$
      & 7.5& 5.1& 11.9& 8.0
      & -7.0& -5.1& 16.7 &11.5
      & 18.8& 12.9& 3.8& 2.8
      & 18.3& 12.7\\
      AudioBERTScore
      & 19.9& 14.0& \underline{28.7}& \underline{19.6}
      & 20.9& 14.2& 22.5& 15.4
      & 9.4& 6.3& 28.1& 19.4
      & 17.8& 12.3\\
      \rowcolor{gray!10}
      \multicolumn{15}{l}{\textit{Reference-free Metrics}}\\
      PAM
      & \underline{22.9}& \underline{15.7}& 17.6& 11.7
      & 18.0& 12.1& 0.6& 0.6
      & \underline{19.7}& \underline{13.3}& 16.0& 10.8
      & 9.2& 6.2\\
      CLAPScore$_{\text{MS}}$
      & 6.9& 4.5& 15.8& 10.6
      & 7.6& 5.1& 27.7& 18.8
      & 8.7& 6.0& 25.9& 17.7
      & 15.2& 10.4\\
      CLAPScore$_{\text{LAION}}$
      & 19.3& 13.6& 22.1& 15.3
      & 11.7& 7.8& 24.7& 17.0
      & 2.6& 1.8& \underline{34.1}& \underline{23.7}
      & 18.3& 12.6\\
      CLAPScore$_{\text{Human}}$
      & 13.6& 9.5& 26.7& 18.7
      & \underline{21.1}& \underline{14.7}& \underline{32.9}& \underline{22.7}
      & 17.8& 12.1& 9.1& 6.2
      & \underline{31.5}& \underline{21.7}\\
      \midrule
      \multirow{2}{*}{ELSA (Ours)}
      & $\bm{33.9}$& $\bm{23.5}$& $\bm{46.5}$& $\bm{32.7}$
      & $\bm{41.2}$& $\bm{28.7}$& $\bm{39.8}$& $\bm{27.5}$
      & $\bm{32.1}$& $\bm{22.4}$& $\bm{36.8}$& $\bm{25.2}$
      & $\bm{37.9}$& $\bm{26.2}$\\
      & (+11.0)& (+7.8)& (+17.8)& (+13.1)
      & (+20.1)& (+14.0)& (+6.9)& (+4.8)
      & (+12.4)& (+9.1)& (+2.7)& (+1.5)
      & (+6.4)& (+4.5)\\
      \bottomrule
    \end{tabular}
    }
    \vspace{-2mm} 
\end{table*}

\vspace{-2mm} 
\subsection{Hierarchical Alignment Scoring}
\vspace{-2mm} 
Given the hierarchical acoustic event representations, we compute the final score by jointly evaluating global text–audio matching and fine-grained alignment of event-level user intent.
We first compute a coarse global matching score $\hat{y}_c$ as the cosine similarity over the entire input as $\hat{y}_c = \bm{t}^\top \bm{a} / \left({\lVert \bm{t} \rVert \, \lVert \bm{a} \rVert}\right)$.
To further evaluate fine-grained alignment at the acoustic-event level, we define the pairwise similarity between event-level text and audio representations for each pair $(i,j)$ as $f_{ij} = \left(\bm{t}_f^{(i)}\right)^\top\bm{a}_f^{(j)}\Big/\left(\lVert \bm{t}_f^{(i)} \rVert\, \lVert \bm{a}_f^{(j)} \rVert\right)$.
Based on these similarities, we obtain event-level precision $P_f$ and recall $R_f$ as
\begin{align}
P_f
&= \frac{1}{M}\sum_{i\in M}
\left(
\max_{j\in M}f_{ji} - \frac{1}{M}\sum_{j\in M} f_{ji}
\right), \\
R_f
&= \frac{1}{M}\sum_{i\in M}
\left(
\max_{j\in M}f_{ij} - \frac{1}{M}\sum_{j\in M} f_{ij}
\right),
\end{align}
which respectively quantify audio-to-text event matching quality and text-to-audio event matching quality.
We then compute the F1-score of $P_f$ and $R_f$, defined as $\hat{y}_f = {2P_fR_f}/ \left(P_f + R_f\right)$, which serves as the fine-grained matching score.
Finally, we adaptively combine both scores to obtain the final evaluation score $\hat{y}$. 
To assign greater weight to fine-grained matching when the number of events is large, we integrate $\hat{y}_c$ and $\hat{y}_f$ as $\hat{y} = \lambda^{M} \hat{y}_c + \left(1 - \lambda^{M}\right) \hat{y}_f$.
Here, $\lambda \in [0,1]$ is a normalization factor that balances coarse- and fine-grained evaluation, and is empirically set to $0.4$ in all experiments. 
\vspace{-2mm} 
\section{Experiments}
\vspace{-2mm} 
\subsection{Datasets and Data Pre-processing}
\vspace{-2mm} 

We evaluated the correlation between the proposed metric and human subjective ratings using four TTA benchmarks: AudioCaps~\cite{AudioCaps}, Clotho~\cite{Clotho}, MusicCaps~\cite{MusicCaps}, and RELATE~\cite{RELATE}.
For each benchmark, audio samples are generated from text queries using multiple TTA models (e.g., AudioLDM~\cite{AudioLDM}) and annotated with human ratings.
As human evaluation criteria, the relation of audio to text caption (REL) is provided across all benchmarks. 
In addition, human perception of quality (OVL) is available for three benchmarks, excluding RELATE. 
REL assesses the semantic consistency between the text query and the generated audio, whereas OVL measures the perceptual quality of the generated audio itself.
We use the human evaluation test sets collected by Deshmukh et al.\footnote{https://github.com/soham97/PAM/tree/main} for AudioCaps and MusicCaps, and the test set collected by Kishi et al.\footnote{https://github.com/lourson1091/audiobertscore} for Clotho.

To further investigate the correlation between the proposed metric and compositional text–audio alignment, we additionally evaluated inclusion of sound events (IS) and order of sound events (OS) scores provided in the RELATE benchmark, as well as CompA-attribute and CompA-order from the CompA~\cite{CompA} benchmark.
In RELATE, IS and OS correspond to human subjective ratings of the coverage and temporal ordering of acoustic events between the text query and the generated audio, respectively.
In CompA, CompA-attribute and CompA-order are evaluated by constructing contrastive text–audio pairs in which the attribute binding and the acoustic event order are deliberately perturbed, respectively, and assessing whether similarity-based metrics correctly identify the matching text–audio pair.

For data preprocessing, all audio samples are resampled to 16 kHz and either cropped or zero-padded to a fixed duration of 10 seconds. 
Subsequently, human subjective ratings are averaged across subjects and are normalized to the range [0, 1] to ensure comparability across benchmarks.

\vspace{-2mm} 
\subsection{Baseline Metrics}
\vspace{-2mm} 
We compared our proposed metric with eight standard automatic evaluation metrics commonly used for TTA generation. 
For reference-based metrics, following Evans et al.~\cite{StableAudio}, we adopted scale-invariant signal-to-distortion ratio (SI-SDR)~\cite{SI-SDR}, Kullback–Leibler divergence in the PaSST~\cite{PaSST} feature space (KL\textsubscript{PaSST}), Fréchet distance in the OpenL3~\cite{Openl3} feature space (FD\textsubscript{OpenL3}), and AudioBERTScore~\cite{AudioBERTScore}. 
As reference-free metrics, which serves as a direct baseline for the proposed metric, we employed PAM~\cite{PAM} and CLAPScore~\cite{CLAPScore}.
For CLAPScore, we evaluated three variants corresponding to different underlying CLAP models: CLAPScore\textsubscript{MS}, CLAPScore\textsubscript{LAION}, and CLAPScore\textsubscript{Human}, based on MS-CLAP~\cite{MS-CLAP}, LAION-CLAP~\cite{LAION-CLAP}, and Human-CLAP~\cite{HumanCLAP}, respectively.
\vspace{-2mm} 
\section{Results and Analysis}
\vspace{-2mm} 
\subsection{Correlation with Human Subjective Ratings}
\vspace{-2mm} 
Table~\ref{tab:quantitative_main} shows the correlation between human subjective ratings and the proposed metric, along with baseline metrics, across four benchmarks (AudioCaps~\cite{AudioCaps}, Clotho~\cite{Clotho}, MusicCaps~\cite{MusicCaps}, and RELATE~\cite{RELATE}).
To enable fair comparison across metrics with heterogeneous output scales, we employed Spearman’s rank correlation coefficient ($\rho$) and Kendall’s rank correlation coefficient ($\tau$) as scale-invariant rank-based correlation measures. 
As shown in the table, the proposed metric consistently exhibited a higher correlation with human subjective ratings than all baseline metrics. 
In particular, for the REL measured by Kendall’s $\tau$, the proposed metric outperformed the best baseline by 13.1, 4.8, 1.5, and 4.5 points on AudioCaps, Clotho, MusicCaps, and RELATE, respectively.
Notably, the proposed metric consistently surpassed CLAPScore\textsubscript{Human}, which computes text–audio similarity within the same feature space as the proposed metric (i.e., Human-CLAP~\cite{HumanCLAP}). 
These results demonstrate the effectiveness of the proposed hierarchical text–audio matching at the acoustic-event level over global similarity matching.

\begin{table}[t]
    \centering
    \caption{Performance comparison on compositional benchmarks with respect to acoustic event attributes and temporal order.
    ``---'' denotes non-executable settings.}
    \vspace{-3mm} 
    \label{tab:event_order} 
    \resizebox{\linewidth}{!}{
    \begin{tabular}{l c c c c c c}
      \toprule
      \multirow{3}{*}{Metrics~[\%]}&\multicolumn{2}{c}{RELATE}& \multicolumn{4}{c}{CompA}\\
      \cmidrule(l{1mm}r{1mm}){2-3}
      \cmidrule(l{1mm}r{1mm}){4-7}
      &IS&OS& \multicolumn{2}{c}
      {Attribute}&\multicolumn{2}{c}{Order}\\
      & $\tau$& $\tau$&text $\uparrow$&audio $\uparrow$&text $\uparrow$&audio $\uparrow$\\
      \midrule
      PAM& 8.8& 5.9& ---& ---& ---& ---\\
      CLAPScore$_{\text{MS}}$& 11.5& 5.3& $\bm{23.9}$& \underline{4.1}& 16.5& 5.5\\
      CLAPScore$_{\text{LAION}}$& 12.9& \underline{10.2}& 19.3& \underline{4.1}& 19.0& 6.0\\
      CLAPScore$_{\text{Human}}$& \underline{20.7}& 10.1&17.3& 3.6& \underline{25.3}& \underline{7.3}\\
      \midrule
      \multirow{2}{*}{ELSA (Ours)}& $\bm{26.7}$& $\bm{13.9}$& \underline{23.4}& $\bm{16.2}$& $\bm{28.0}$& $\bm{15.5}$\\
      &(+6.0)& (+3.7)& (-0.5)& (+12.1)& (+2.7)& (+8.2)\\
      \bottomrule
    \end{tabular}
    }
    \vspace{-6mm}
\end{table}
Table~\ref{tab:event_order} presents the correlation between the proposed metric and text–audio compositional similarity on the RELATE~\cite{RELATE} and CompA~\cite{CompA} benchmarks.
For RELATE, we evaluated the correlation on IS and OS using Kendall’s $\tau$.
For CompA, we evaluated both CompA-attribute and CompA-order tasks in terms of retrieval accuracy, considering text retrieval from audio (“text” in the table) and audio retrieval from text (“audio” in the table).
As shown in Table~\ref{tab:event_order}, the proposed metric consistently outperformed baseline metrics on RELATE and achieved superior performance on three out of four evaluation settings on CompA.
These results suggest that explicitly extracting acoustic event–level representations improves alignment with human subjective ratings in compositional aspects, beyond overall text-audio relevance.

\vspace{-2mm} 
\subsection{Component Ablation Studies}
\vspace{-2mm} 
\begin{table}[t]
    \centering
    \caption{Ablation study of the LASS model and text-audio embedding feature spaces on the Clotho benchmark.}
    \label{tab:ablation} 
    \vspace{-3mm} 
    \resizebox{\linewidth}{!}{
    \begin{tabular}{c c c c c c c c c c c c c }
      \toprule
      {[\%]}&
      \multirow{2}{*}{\makecell{LASS\\Model}}&
      \multirow{2}{*}{\makecell{Feature\\Space}}&\multicolumn{2}{c}{OVL}& \multicolumn{2}{c}{REL}\\
      Metrics&&&$\rho$& $\tau$&$\rho$& $\tau$\\
      \midrule
      (i)& SAM Audio& Human-CLAP&$\bm{41.2}$& $\bm{28.7}$& 39.8& 27.5\\
      \midrule
      (ii)& AudioSep& Human-CLAP& 40.9& 28.3& $\bm{43.9}$& $\bm{30.6}$\\
      (iii)& SoloAudio& Human-CLAP& 37.1& 25.9& 36.5& 25.2\\
      (iv)& SAM Audio& MS-CLAP& 28.2& 19.2& 26.4& 17.9\\
      (v)& SAM Audio& LAION-CLAP& 31.5& 21.5& 30.4& 20.6\\
      \bottomrule
    \end{tabular}
    \vspace{-2mm}
    }
\end{table}
To investigate the effectiveness of the proposed metric, we conducted ablation studies on its major components using the Clotho~\cite{Clotho} benchmark.
Table~\ref{tab:ablation} summarizes the impact on correlation with human subjective ratings when varying the LASS model and the feature space used for text–audio similarity comparison.
For the LASS model, we compared (i) SAM-Audio~\cite{SAM-Audio} adopted in the proposed metric with (ii) AudioSep~\cite{AudioSep} and (iii) SoloAudio~\cite{SoloAudio}.
For the text–audio embedding feature space, we replaced (i) Human-CLAP with (iv) MS-CLAP~\cite{MS-CLAP} and (v) LAION-CLAP~\cite{LAION-CLAP}.

As shown in Table~\ref{tab:ablation}, the performance differences induced by altering the LASS model were relatively limited: in terms of Kendall’s $\tau$, the maximum gap was 2.8 points for OVL across Metrics (i)–(iii), and 5.4 points for REL between Metrics (ii) and (iii). 
In contrast, altering the feature space for text–audio comparison led to substantially larger differences, with maximum gaps of 9.5 points for OVL and 9.6 points for REL when comparing Metrics (i) and (iv), respectively.
These results empirically indicate that the proposed metric is considerably more sensitive to the choice of text–audio embedding space than to the LASS model employed.

\begin{figure}[t]
    \centering
    \includegraphics[width=\linewidth]{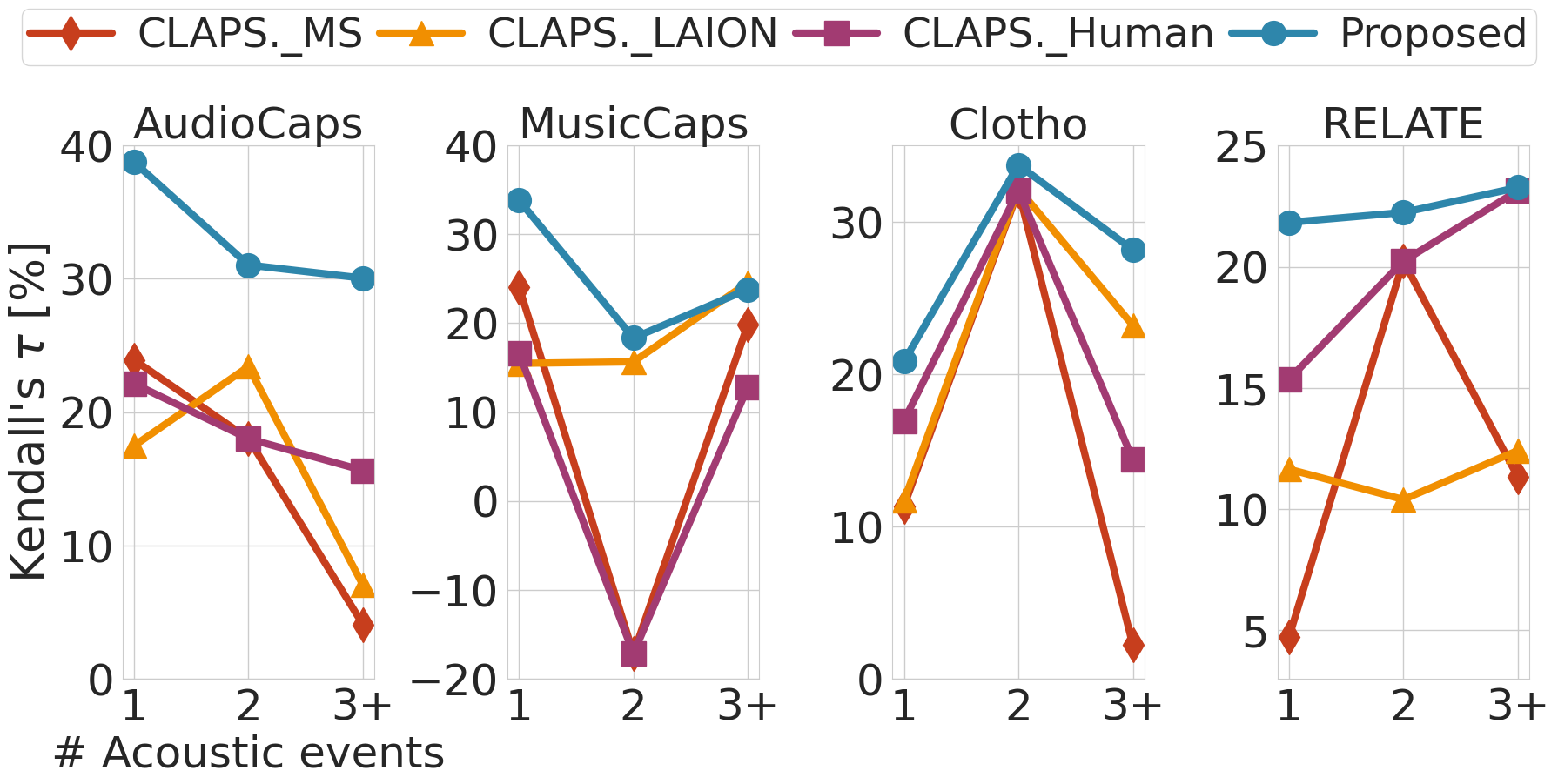}
    \vspace{-3mm} 
    \caption{Sensitivity analysis of metric–REL correlation with respect to the number of acoustic events in the text query.}
    \label{fig:sens_analysis}
    \vspace{-2mm}
\end{figure}

\subsection{Sensitivity to the Complexity of User Intent}
\vspace{-2mm} 
Figure~\ref{fig:sens_analysis} evaluates how the proposed metric’s correlation with human subjective ratings varies with the number of acoustic events in each query, thereby evaluating its robustness to increasing user-intent complexity.
Here, we used the text-only LLM (GPT-5.2) to estimate the number of events per query and categorized them into three groups: one, two, and three or more events. 
We then compare the proposed metric with CLAPScore\textsubscript{MS}, CLAPScore\textsubscript{LAION}, and CLAPScore\textsubscript{Human} in terms of Kendall’s $\tau$ on REL.

As shown in Figure~\ref{fig:sens_analysis}, the proposed metric consistently outperformed all baseline metrics across all acoustic-event conditions on four benchmarks.
In particular, on the RELATE benchmark, the maximum variation across event-count conditions is 1.4 points for the proposed metric, compared to 15.5, 2.0, and 7.8 points for CLAPScore\textsubscript{MS}, CLAPScore\textsubscript{LAION}, and CLAPScore\textsubscript{Human}, respectively.
These results suggest that the proposed metric exhibits a high correlation with human subjective ratings regardless of the number of events in the text query, indicating robustness to variations in user-intent complexity.

\subsection{Analysis of Score Distributions}
\begin{figure}[t]
    \centering
    \includegraphics[width=\linewidth]{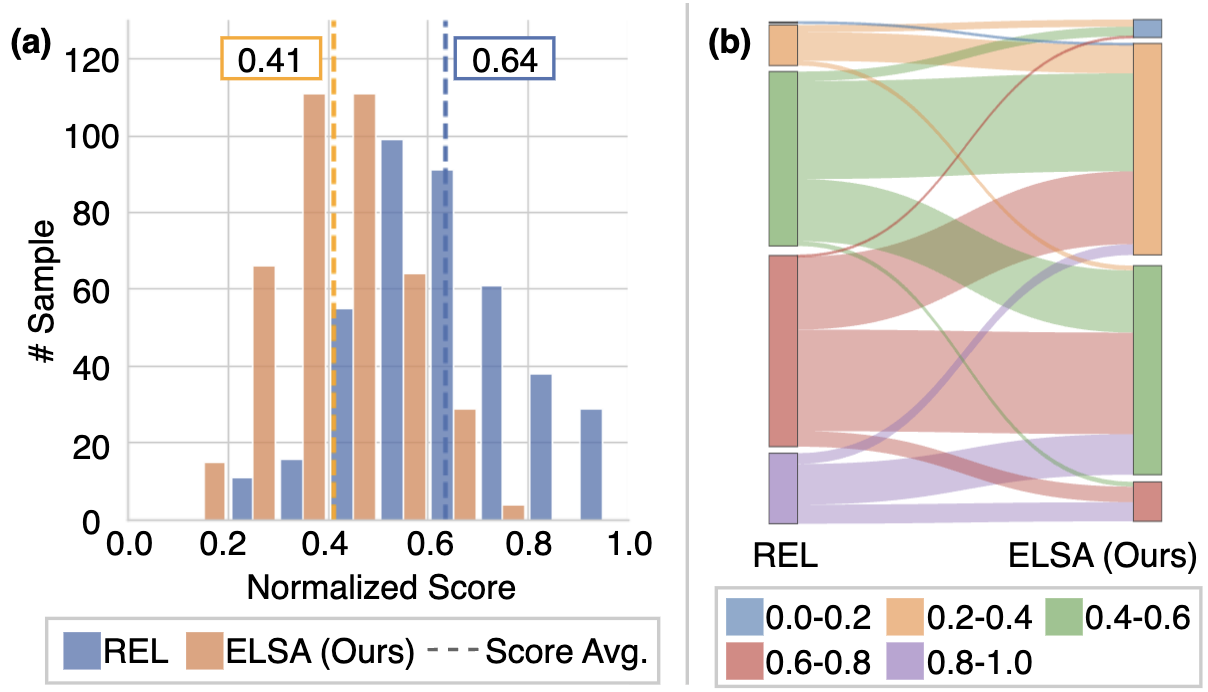}
    \vspace{-4mm}
    \caption{Relationship between REL and ELSA on AudioCaps. (a) Histogram of normalized score distributions. (b) Sankey diagram visualizing score correspondence with a bin width of 0.2.}
    \label{fig:score_dist}
    \vspace{-2mm}
\end{figure}

\vspace{-2mm}
To obtain a detailed interpretation of the proposed metric, Figure~\ref{fig:score_dist} illustrates the relationship between REL and ELSA on the AudioCaps benchmark. 
Figure~\ref{fig:score_dist}-(a) shows histograms of the normalized score distributions for REL and ELSA, while Figure~\ref{fig:score_dist}-(b) depicts their correspondence using a Sankey diagram with a bin width of 0.2.
As shown in Figure~\ref{fig:score_dist}-(a), both REL and ELSA follow Gaussian-like distributions.
However, their mean values differ: the average scores are 0.64 for REL and 0.41 for ELSA, indicating that ELSA scores are lower by 0.23 points on average.
Furthermore, Figure~\ref{fig:score_dist}-(b) demonstrates a systematic shift between the two scales. 
Each REL bin primarily corresponds to an ELSA bin that is lower by 0.2 points. 
For instance, the strongest correspondence for the 0.4–0.6 REL bin (green) and the 0.6–0.8 REL bin (red) is observed in the 0.2–0.4 ELSA bin (orange) and the 0.4–0.6 ELSA bin (green), respectively.

Taken together, these findings indicate that while ELSA closely approximates the distributional shape of human ratings, its absolute scoring scale is systematically shifted relative to REL. 
Accordingly, although the proposed metric is effective for evaluating TTA systems and for developing human preference-optimized TTA methods~\cite{Tango2, TangoFlux}, further calibration of its output scale remains an avenue for future improvement.

\section{Limitation}
\vspace{-2mm} 
As a primary limitation, ELSA does not explicitly model the temporal order of acoustic events.
Although ELSA outperforms baseline metrics on order-sensitive benchmarks, such as OS on RELATE~\cite{RELATE} and CompA-order on CompA~\cite{CompA} (Table~\ref{tab:event_order}), explicitly modeling event duration and sequential structure remains a promising direction for future improvement.
\section{Conclusion}
\vspace{-2mm} 
In this paper, we proposed ELSA, a reference-free evaluation metric for TTA generation that enables fine-grained text–audio comparison.
Experimental results show that ELSA consistently outperforms existing metrics, including both reference-based and reference-free approaches.
Furthermore, ablation and sensitivity analyses show both component-wise effectiveness and robustness to variations in the number of target acoustic events.

\newpage
\section{Acknowledgments}
{
Part of this study was executed in the ``Honda Bridge Project,'' a collaborative research and education program between the Faculty of Science and Technology at Keio University and Honda Motor Co., Ltd.
}

\section{Generative AI Use Disclosure}
Generative AIs were used solely for auxiliary purposes, such as language refinement, manuscript formatting, and the implementation of standard algorithms.
They were not involved in the research design, nor did they contribute to the development, implementation, or evaluation of the proposed methods.
Accordingly, Generative AI did not constitute an essential, original, or non-standard component of this research.

\bibliographystyle{IEEEtran}
\bibliography{mybib}

@inproceedings{PAM,
  title     = {{PAM: Prompting Audio-Language Models for Audio Quality Assessment}},
  author    = {Soham Deshmukh and Dareen Alharthi and Benjamin Elizalde and Hannes Gamper and Mahmoud {Al Ismail} and Rita Singh and Bhiksha Raj and Huaming Wang},
  year      = {2024},
  booktitle = {{Interspeech}},
  pages     = {3320--3324},
  doi       = {10.21437/Interspeech.2024-325},
  issn      = {2958-1796},
}

@InProceedings{AudioLDM,
  title = 	 {{A}udio{LDM}: Text-to-Audio Generation with Latent Diffusion Models},
  author =       {Liu, Haohe and Chen, Zehua and Yuan, Yi and Mei, Xinhao and Liu, Xubo and Mandic, Danilo and Wang, Wenwu and Plumbley, Mark D},
  booktitle = 	 {ICML},
  pages = 	 {21450--21474},
  year = 	 {2023},
  volume = 	 {202},
  month = 	 {23--29 Jul},
}

@inproceedings{RELATE,
  title     = {{RELATE: Subjective evaluation dataset for automatic evaluation of relevance between text and audio }},
  author    = {Yusuke Kanamori and Yuki Okamoto and Taisei Takano and Shinnosuke Takamichi and Yuki Saito and Hiroshi Saruwatari},
  year      = {2025},
  booktitle = {{Interspeech}},
  pages     = {3155--3159},
  doi       = {10.21437/Interspeech.2025-1830},
  issn      = {2958-1796},
}

@inproceedings{AudioBERTScore,
  author    = {Minoru Kishi and Ryosuke Sakai and Shinnosuke Takamichi and Yusuke Kanamori and Yuki Okamoto},
  title     = {{AudioBERTScore: Objective Evaluation of Environmental Sound Synthesis Based on Similarity of Audio Embedding Sequences}},
  booktitle = {AAAI Audio-Centric AI Workshop},
  year      = {2026},
}

@INPROCEEDINGS{Clotho,
  author={Drossos, Konstantinos and Lipping, Samuel and Virtanen, Tuomas},
  booktitle={ICASSP}, 
  title={{Clotho: an Audio Captioning Dataset}}, 
  year={2020},
  volume={},
  number={},
  pages={736-740},
}

@inproceedings{AudioCaps,
    author    = {Chris Dongjoo Kim and Byeongchang Kim and Hyunmin Lee and Gunhee Kim},
    title     = {{AudioCaps: Generating Captions for Audios in The Wild}},
    booktitle = {NAACL-HLT},
    year      = 2019
}

@article{MusicCaps,
  title={{MusicLM: Generating Music From Text}},
  author={Agostinelli, Andrea and Denk, Timo I and Borsos, Zal{\'a}n and Engel, Jesse and Verzetti, Mauro and Caillon, Antoine and Huang, Qingqing and Jansen, Aren and Roberts, Adam and Tagliasacchi, Marco and others},
  journal={arXiv:2301.11325},
  year={2023}
}

@inproceedings{StableAudio,
  title={{Fast Timing-Conditioned Latent Audio Diffusion}},
  author={Evans, Zach and Carr, CJ and Taylor, Josiah and Hawley, Scott H and Pons, Jordi},
  booktitle={ICML},
  year={2024}
}

@INPROCEEDINGS{HumanCLAP,
  author={Takano, Taisei and Okamoto, Yuki and Kanamori, Yusuke and Saito, Yuki and Nagase, Ryotaro and Saruwatari, Hiroshi},
  booktitle={APSIPA ASC}, 
  title={{Human-CLAP: Human-perception-based Contrastive Language-audio Pretraining}}, 
  year={2025},
  volume={},
  number={},
  pages={131-136},
}

@INPROCEEDINGS{LAION-CLAP,
  author={Wu, Yusong and Chen, Ke and Zhang, Tianyu and Hui, Yuchen and Berg-Kirkpatrick, Taylor and Dubnov, Shlomo},
  booktitle={ICASSP}, 
  title={{Large-Scale Contrastive Language-Audio Pretraining with Feature Fusion and Keyword-to-Caption Augmentation}}, 
  year={2023},
  volume={},
  number={},
  pages={1-5},
}

@inproceedings{CLAPScore,
  title={{A Reference-free Metric for Language-Queried Audio Source Separation using Contrastive Language-Audio Pretraining}},
  author={Xiao, Feiyang and Guan, Jian and Zhu, Qiaoxi and Liu, Xubo and others},
  booktitle={DCASE Workshop},
  year={2024}
}

@INPROCEEDINGS{MS-CLAP,
  author={Elizalde, Benjamin and Deshmukh, Soham and Ismail, Mahmoud Al and Wang, Huaming},
  booktitle={ICASSP}, 
  title={{CLAP Learning Audio Concepts from Natural Language Supervision}}, 
  year={2023},
  volume={},
  number={},
  pages={1-5},
}

@INPROCEEDINGS{SI-SDR,
  author={Roux, Jonathan Le and Wisdom, Scott and Erdogan, Hakan and Hershey, John R.},
  booktitle={ICASSP}, 
  title={{SDR – Half-baked or Well Done?}}, 
  year={2019},
  volume={},
  number={},
  pages={626-630},
}

@INPROCEEDINGS{Openl3,
  author={Cramer, Aurora Linh and Wu, Ho-Hsiang and Salamon, Justin and Bello, Juan Pablo},
  booktitle={ICASSP 2019}, 
  title={{Look, Listen, and Learn More: Design Choices for Deep Audio Embeddings}}, 
  year={2019},
  volume={},
  number={},
  pages={3852-3856},
}

@inproceedings{PaSST,
  title     = {{Efficient Training of Audio Transformers with Patchout}},
  author    = {{Khaled Koutini and Jan Schlüter and Hamid Eghbal-zadeh and Gerhard Widmer}},
  year      = {{2022}},
  booktitle = {{Interspeech}},
  pages     = {{2753--2757}},
}

@inproceedings{AudioGen,
title={{AudioGen: Textually Guided Audio Generation}},
author={Felix Kreuk and Gabriel Synnaeve and Adam Polyak and Uriel Singer and Alexandre D{\'e}fossez and Jade Copet and Devi Parikh and Yaniv Taigman and Yossi Adi},
booktitle={ICLR},
year={2023},
}

@article{AudioLDM2,
  title={{AudioLDM 2: Learning Holistic Audio Generation With Self-Supervised Pretraining}},
  author={Liu, Haohe and Yuan, Yi and Liu, Xubo and Mei, Xinhao and Kong, Qiuqiang and Tian, Qiao and Wang, Yuping and others},
  journal={TASLP},
  volume={32},
  pages={2871--2883},
  year={2024},
}

@INPROCEEDINGS{PixelSNAIL,
  author={Liu, Xubo and Iqbal, Turab and Zhao, Jinzheng and Huang, Qiushi and others},
  booktitle={MLSP}, 
  title={{Conditional Sound Generation Using Neural Discrete Time-Frequency Representation Learning}}, 
  year={2021},
  volume={},
  number={},
  pages={1-6},
}

@inproceedings{DAG,
  title={{Full-band General Audio Synthesis with Score-based Diffusion}},
  author={Pascual, Santiago and Bhattacharya, Gautam and Yeh, Chunghsin and Pons, Jordi and Serr{\`a}, Joan},
  booktitle={ICASSP},
  pages={1--5},
  year={2023},
}

@inproceedings{lloyd,
  title={{Sound synthesis for impact sounds in video games}},
  author={Lloyd, D Brandon and Raghuvanshi, Nikunj and Govindaraju, Naga K},
  booktitle={I3D},
  pages={55--62},
  year={2011}
}

@misc{riffusion,
  author       = {Forsgren, Seth and Martiros, Hayk},
  title        = {{Riffusion: Stable Diffusion for Real-Time Music Generation}},
  year         = {2022},
  howpublished = {URL: \url{https://riffusion.com/about}},
}

@inproceedings{HICEScore, author = {Zeng, Zequn and Sun, Jianqiao and Zhang, Hao and Wen, Tiansheng and Su, Yudi and Xie, Yan and Wang, Zhengjue and Chen, Bo}, 
title = {{HICEScore: A Hierarchical Metric for Image Captioning Evaluation}}, 
year = {2024}, 
booktitle = {ACM-MM}, pages = {866–875}, numpages = {10}
}

@inproceedings{EMScore,
  author    = {Yaya Shi and
               Xu Yang and
               Haiyang Xu and
               Chunfeng Yuan and
               Bing Li and
               Weiming Hu and
               Zheng{-}Jun Zha},
  title     = {{EMScore: Evaluating Video Captioning via Coarse-Grained and Fine-Grained
               Embedding Matching}},
  booktitle = {CVPR},
  year      = {2022},
}

@article{SAM-Audio,
    title={{SAM Audio: Segment Anything in Audio}},
    author={Bowen Shi and Andros Tjandra and John Hoffman and Helin Wang and Yi-Chiao Wu and Luya Gao and Julius Richter and Matt Le and Apoorv Vyas and Sanyuan Chen and others},
    year={2025},
    journal={arXiv preprint arXiv:2512.18099},
}

@inproceedings{
CompA,
title={{CompA: Addressing the Gap in Compositional Reasoning in Audio-Language Models}},
author={Sreyan Ghosh and Ashish Seth and Sonal Kumar and Utkarsh Tyagi and Chandra Kiran Reddy Evuru and Ramaneswaran S and S Sakshi and Oriol Nieto and others},
booktitle={ICLR},
year={2024},
}

@article{AudioSep,
  title={{Separate Anything You Describe}},
  author={Liu, Xubo and Kong, Qiuqiang and Zhao, Yan and Liu, Haohe and Yuan, Yi and Liu, Yuzhuo and Xia, Rui and Wang, Yuxuan and Plumbley, Mark D and Wang, Wenwu},
  journal={TASLP},
  year={2024},
}

@inproceedings{SoloAudio,
  title={{SoloAudio: Target Sound Extraction with Language-oriented
Audio Diffusion Transformer}},
  author={Wang, Helin and Hai, Jiarui and Lu, Yen-Ju and Thakkar, Karan and Elhilali, Mounya and Dehak, Najim},
  booktitle={ICASSP},
  pages={1--5},
  year={2025},
}

@article{lan,
  title={{A Survey of Automatic Evaluation Methods on Text, Visual and Speech Generations}},
  author={Lan, Tian and Zhou, Yang-Hao and Ma, Zi-Ao and Sun, Fanshu and Sun, Rui-Qing and Luo, Junyu and others},
  journal={arXiv preprint arXiv:2506.10019},
  year={2025}
}

@article{su,
  title={{Audio-Language Models for Audio-Centric Tasks:
A survey}},
  author={Su, Yi and Bai, Jisheng and Xu, Qisheng and Xu, Kele and Dou, Yong},
  journal={arXiv preprint arXiv:2501.15177},
  year={2025}
}

@ARTICLE{SDR,
  author={Vincent, E. and Gribonval, R. and Fevotte, C.},
  journal={TASLP}, 
  title={{Performance measurement in blind audio source separation}}, 
  year={2006},
  volume={14},
  number={4},
  pages={1462-1469},
}

@inproceedings{vela,
    title = {{VELA}: An {LLM}-Hybrid-as-a-Judge Approach for Evaluating Long Image Captions},
    author = {Matsuda, Kazuki and Wada, Yuiga and Hirano, Shinnosuke and Otsuki, Seitaro and Sugiura, Komei},
    booktitle = {EMNLP},
    year = {2025},
    pages = {8680--8696},
}

@inproceedings{pearl,
    title = {{LLM-Free Image Captioning Evaluation in Reference-Flexible Settings}},
    author = {Hirano, Shinnosuke  and Wada, Yuiga and Matsuda, Kazuki and Otsuki, Seitaro and  Sugiura, Komei},
    booktitle = {AAAI},
    year = {2026},
}

@inproceedings{Tango2,
  title={{Tango 2: Aligning diffusion-based text-to-audio generations through direct preference optimization}},
  author={Majumder, Navonil and Hung, Chia-Yu and Ghosal, Deepanway and Hsu, Wei-Ning and others},
  booktitle={ACM MM},
  pages={564--572},
  year={2024}
}

@inproceedings{TangoFlux,
title={{TangoFlux: Text to Audio Generation with {CLAP}-Ranked Preference Optimization}},
author={Chia-Yu Hung and Navonil Majumder and Zhifeng Kong and Ambuj Mehrish and Amir Zadeh and Chuan Li and Rafael Valle and Bryan Catanzaro and Soujanya Poria},
booktitle={ICLR},
year={2026},
}

@inproceedings{AudioX,
  title={{AudioX: Diffusion Transformer for Anything-to-Audio Generation}},
  author={Tian, Zeyue and Jin, Yizhu and Liu, Zhaoyang and Yuan, Ruibin and Tan, Xu and Chen, Qifeng and Xue, Wei and Guo, Yike},
  booktitle={ICLR},
  year={2026}
}

@inproceedings{MGA-CLAP,
  title={Advancing multi-grained alignment for contrastive language-audio pre-training},
  author={Li, Yiming and Guo, Zhifang and Wang, Xiangdong and Liu, Hong},
  booktitle={ACM MM},
  pages={7356--7365},
  year={2024}
}

@article{FineLAP,
  title={FineLAP: Taming Heterogeneous Supervision for Fine-grained Language-Audio Pretraining},
  author={Li, Xiquan and Xu, Xuenan and Ma, Ziyang and Chen, Wenxi and He, Haolin and Kong, Qiuqiang and Chen, Xie},
  journal={ACL},
  year={2026}
}

\end{document}